# Centrality Dependent Lévy Analysis of two-pion BEC Functions at PHENIX

*Tamás* Novák[1,2*] *for the PHENIX Collaboration*

[1] MATE KRC, H-3200 Gyöngyös, Mátrai út 36, Hungary
[2] BGE, H-1054 Budapest, Alkotmány út 9-11, Hungary

**Abstract.** We present most recent PHENIX preliminary data on centrality dependence of two-pion Bose-Einstein correlation functions measured in $\sqrt{s_{NN}} = 200$ GeV Au+Au collisions at the Relativistic Heavy Ion Collider (RHIC). The data are well described by assuming the source to be a Lévy-stable distribution. The Lévy parameters, $R$, $\alpha$, $\lambda$ are measured in 18 bins of transverse mass ($m_T$) for 4 centrality intervals. The Lévy scale parameter $R(m_T)$ decreases with $m_T$ and exhibits a clear centrality ordering which supports its geometrical interpretation. The Lévy exponent $\alpha(m_T)$ is independent of $m_T$ in every centrality bin but shows some centrality dependence. At all centralities $\alpha$ is significantly different from that of a Gaussian ($\alpha = 2$) or Cauchy ($\alpha = 1$) source distribution. The Lévy strength parameter $\lambda$ was also investigated. We observed that $\lambda(m_T)$ decreases at low $m_T$ and saturates around $m_T = 0.6$ GeV/$c^2$. In the already published 0-30% centrality class, Monte Carlo simulations are found to be inconsistent with the measurements, unless a significant reduction of the in-medium mass of the $\eta'$ meson is included.

## 1 Introduction

In particle and nuclear physics intensity interferometry provides a direct experimental method for the determination of sizes, shapes and lifetimes of particle-emitting sources (for recent reviews see [1, 2]). In particular, boson interferometry provides a powerful tool for the investigation of the space-time structure of particle production processes, since Bose-Einstein correlations (BEC) of two identical bosons reflect both geometrical and dynamical properties of the particle radiating source.

The technique was discovered in astrophysics by R. Hanbury Brown and Q. R. Twiss in correlation measurements, that were performed in radio and optical astronomy to measure the angular diameters of stars [3]. Hanbury Brown and Twiss are considered the experimental founders of the HBT effect. Independently, the intensity correlations of identical pions were observed in proton-antiproton annihilation. These correlations were explained by G. Goldhaber, S. Goldhaber, W-Y. Lee and A. Pais [4] on the basis of the Bose-Einstein symmetrization of the wave-function of identical pion pairs.

If interactions between the created hadrons, higher order correlations, decays and all other dynamical two-particle correlations may be neglected, then the two-particle Bose-Einstein correlation function is simply related to the source function $S(x, k)$ (which describes the probability density of particle creation at the space-time point $x$ and with four-momentum $p$).

## 2 The PHENIX Detector

The PHENIX experiment at RHIC was designed to study various type of particles produced in heavy ion collisions, including photons, electrons, muons and charged hadrons.

PHENIX has two central arm spectrometers (East and West), each covering $|\eta| < 0.35$ in pseudorapidity and $\Delta\varphi = \pi/2$ in azimuth. In each central arm, charged particle tracks are reconstructed using hit information from the drift chamber (DC), the first layer of pad chambers (PC1) and the collision $z$-vertex position measured by the beam-beam counters (BBC).

The charged pion identification was done by lead scintillator Electromagnetic Calorimeter (PbSc) as well as the high resolution time-of-flight detectors (TOF East and TOF West). We identified charged pions in 0.2 GeV/$c \leq p_T \leq 0.85$ GeV/$c$ transverse momentum range.

The detailed description of the basic experimental configuration (without the upgrades made after early 2000s) can be found here [5].

## 3 Parametrizations of BEC

The Bose-Einstein correlation function, $C_2$, is measured by $C_2(Q) = \rho(Q)/\rho_0(Q)$, where $\rho(Q)$ is the density of identical boson pairs with invariant four-momentum difference $Q = \sqrt{-(p_1 - p_2)^2}$ and $\rho_0(Q)$ is the similar density in an artificially constructed reference sample, which should differ from the data only in that it does not

---

* Corresponding author: novak.tamas@uni-bge.hu

contain the effects of Bose symmetrization of identical bosons. It is often parametrized as

$$C_2 = \gamma[1 + \lambda G](1 + \varepsilon Q) \quad (1)$$

with

$$G = \exp(-(RQ)^2). \quad (2)$$

The source is frequently assumed to be Gaussian which leads to Gaussian correlation function. So, the corresponding distribution of boson emission points in space-time is a spherically symmetric Gaussian with standard deviation $R$. However, there is no good reason to assume Gaussian source function – except simplicity. Recently many measurements suggested phenomena beyond Gaussian distribution in heavy ion collisions.

Past measurements of two-pion Bose-Einstein correlation functions went beyond the Gaussian approximation show that the precise shape of BEC is indeed not Gaussian. For example, L3 experiment applied Lévy stable distributions first time [6]. The PHENIX experiment also measured correlation functions based on Lévy [7]. So, the best way to depart from the Gaussian is the generalization to a symmetric Lévy stable distribution. It yields to

$$G = \exp(-(RQ)^\alpha), \quad (3)$$

where $0 < \alpha \leq 2$ is the so-called index of stability which was introduced to BEC in [8]. In the case of $\alpha = 2$ the Gaussian case is restored, the case of $\alpha = 1$ corresponds to the Cauchy distribution.

There can be many reasons for the appereance of Lévy distributions: criticality, QCD jets, anomalous diffusion and other phenomena. Lévy-type of analyses were discussed before in Refs. [9 – 11].

It is also important to mention the final state Coulomb-correction which is a crucial part of any correlation measurement dealing with charged particles. One cannot fit the above functional form to the measured correlation functions before properly taking the final state Coulomb repulsion of the identically charged pions into account. For an exhaustive description of this, in particular the functions used in fitting our data, see Refs. [12 – 14].

The Lévy parametrization (incorporating the effect of the final-state Coulomb interactions) is found to give a statistically acceptable description of the data with Lévy $R$, $\alpha$ and $\lambda$ as it can be seen on Fig. 2.

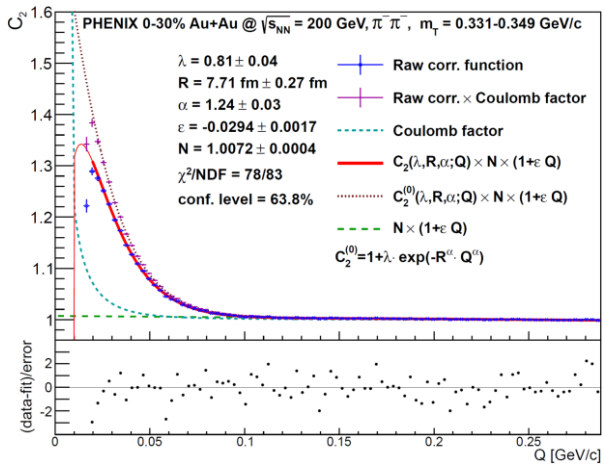

**Fig. 2.** Example fit of Bose-Einstein correlation function.

### 3.1 BEC at the second order QCD phase transition

According to recent lattice QCD calculations the line of the first order phase transition ends at the critical endpoint (CEP), where the transition from hadron gas to quark-gluon-plasma (QGP) becomes a second order phase transition.

In case of second order phase transition the $\eta$ exponent describes the power-law behavior of the spatial correlations at the critical point: $\sim r^{-(d-2+\eta)}$, where $d$ is the dimension [15, 16]. For the three-dimensional Lévy stable sources the correlation between the initial and actual positions decays also as a power-law, where the exponent is given by the Lévy index of stability: $\sim r^{-(1+\alpha)}$ [8, 16]. From comparing these exponents, one can easily see that the Lévy exponent is identical to $\eta$ at the critical point.

Rieger determined the correlation exponent using 3d Ising model [15]: $\alpha(\text{Lévy}) = \eta(\text{3d Ising}) = 0.50 \pm 0.05$. If one extracts the value of the Lévy exponent $\alpha$ at different center-of-mass collision energies, the data may yield information on the nature of the quark-hadron phase transition, particularly it may shed light on the location of the critical endpoint on the phase-diagram.

### 3.2 The shape parameters of BEC

The most important physical parameters $(R, \alpha, \lambda)$ are measured in this work as a function of pair transverse mass $(m_\text{T})$. The transverse mass is defined with $m_\text{T} = \sqrt{m^2 + p_\text{T}^2}$. The $R$ is the length of homogeneity (measures the size), $\alpha$ is the index of stability and $\lambda$ is the so-called impact parameter (responsible for instance particle creation mechanism, like core-halo model [17, 18]).

The effect of the parameters can be seen in Fig. 3.

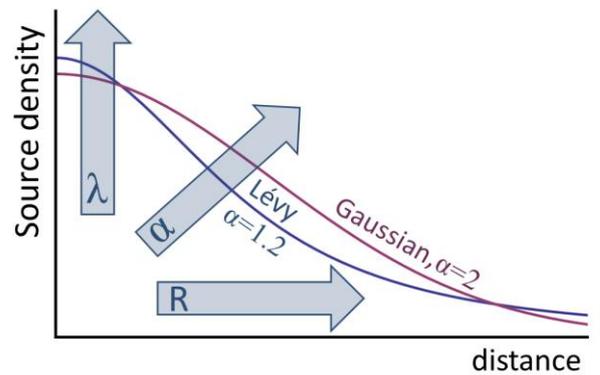

**Fig. 3.** Comparison of the Gaussian and the Lévy distribution with a given $R$, $\alpha$ and $\lambda$. It can be seen that the Lévy distribution has a power-law-like tail.

If there is no halo part and the source is chaotic and fully thermal then $\lambda = 1$. If $\lambda < 1$ it is the sign of the presence of the decay mesons. The $\alpha$ shape parameter or Lévy-index indicates the deviation from the Gaussian case. Moreover, it could be associated – as it was mentioned before – to one of the critical exponents namely to the critical exponent of the spatial correlation in 3-dimension [15]. The $R$ is frequently referred as the Lévy scale parameter.

## 4 Results

We studied the Lévy-parameters in 10% wide centrality bins in 0–40% range and in 18 transverse mass ($m_T$) bins at $\sqrt{s_{NN}} = 200$ GeV collision energy in Au+Au system. We determine the $m_T$ dependencies of the Lévy parameters.

### 4.1 The Lévy scale parameter $R$

First, we measured the $R(m_T)$ dependencies in the above mentioned centrality ranges. The results can be seen in Fig. 4.

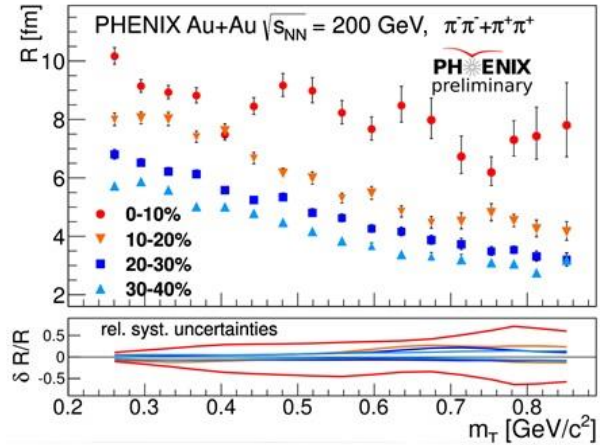

**Fig. 4.** The Lévy scale parameter $R$ as a function of $m_T$ and centrality.

The Lévy scale exhibits decreasing trends as a function of $m_T$, and one can clearly observe a geometrical centrality dependence. In more central collisions, the $R$ parameter (which is related to the physical size of the system) takes larger values.

### 4.2 The Lévy shape parameter $\alpha$

In case of the Lévy exponent $\alpha$ we can see that for all centrality and $m_T$ ranges the values are between 1 and 2. The measured values are far from the Gaussian $\alpha = 2$ case and also far from the conjectured $\alpha \leq 0.5$ at the critical point.

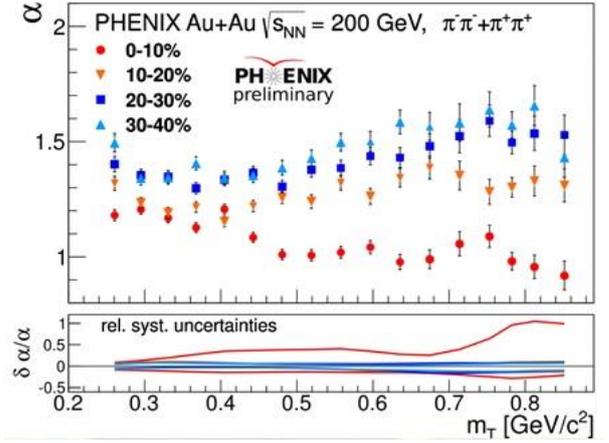

**Fig. 5.** The Lévy shape parameter $\alpha$ as a function of $m_T$ and centrality.

### 4.3 The Lévy strength parameter $\lambda$

As it was mentioned above the $\lambda$ parameter could be related to the core-halo ratio. $\lambda(m_T)$ is shown in Fig. 6. As one can see from the figure, the strength of the correlation function is not equal to unity, and not even constant as a function of $m_T$, the reason for which may be the fact that a large fraction of low $m_T$ pions are produced from decay of long-lived resonances.

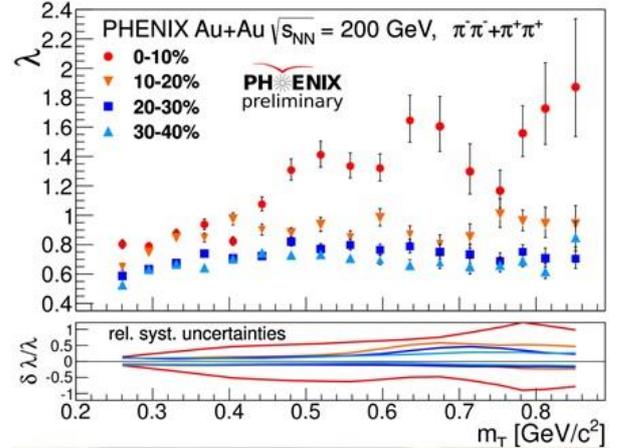

**Fig. 6.** The Lévy strength parameter $\lambda$ as a function of $m_T$ and centrality.

As seen in Fig. 6. $\lambda$ appears to increase with $m_T$ until it saturates around $m_T = 0.6$ GeV/$c^2$. To further study the dependence of $\lambda$ on $m_T$ it is advantageous to use the ratio $\lambda/\lambda_{max}$ where $\lambda_{max}$ is the saturated value of $\lambda$ which we determine in the region $m_T > 0.55$ GeV/$c^2$. Fig. 7. shows the resulting $\lambda/\lambda_{max}$ dependence on $m_T$.

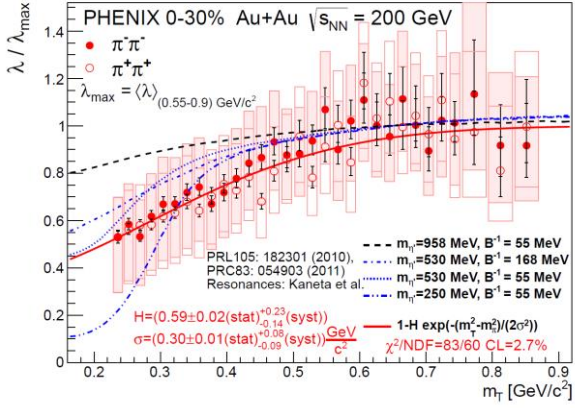

**Fig. 7.** Normalized correlation strength parameter $\lambda/\lambda_{\max}$ versus average $m_T$.

The decreasing trend at lower $m_T$ values can clearly be observed in the case of the $\lambda(m_T)$, but it is clearer in the case of $\lambda(m_T)/\lambda_{\max}$. It has been suggested [19] that $U_A(1)$ symmetry restoration and its related in-medium mass reduction of the $\eta'$ meson in hot, dense hadronic matter would cause a reduction in the value of $\lambda$ at low $m_T$. In Fig. 7, our data are compared with parameter scans from Refs. [20, 21], using different values for the in-medium $\eta'$ mass $m^*_{\eta'}$ and the $\eta'$ condensate temperature (slope parameter) $B^{-1}_{\eta'}$. Our data are seen to be suppressed compared to the prediction with no in-medium $\eta'$ mass modification. Within systematics, our data are not inconsistent with selected parameter scan results of Refs. [20, 21] using a modified in-medium $\eta'$ mass. This phenomenon can be explained in more detail in Refs. [2, 7]. Our data allows for a scrutiny of several theoretical predictions for the in-medium mass modification of $\eta'$ [22 – 27].

It is worth noting that this effect was not observed in the S+Pb data on $\lambda(m_T)$ at $E_{LAB}$ = 200 AGeV CERN SPS energies [28]. Similarly, a result consistent with lack of suppression was reported recently by the NA61 collaboration for Be+Be collisions at $E_{LAB}$ = 150 AGeV CERN SPS energies [29].

## 5 Conclusions

We measured correlation functions in Au+Au collisions at $\sqrt{s_{NN}}$ = 200 GeV colliding energies in the PHENIX experiment in 18 $m_T$ and 4 centrality range. We fit the measured data with Lévy distribution, yield the Lévy parameters and determine their $m_T$ and centrality dependencies.

The Lévy scale parameter $R$ has decreasing trends in all observed centrality bin as a function of $m_T$ and clearly shows geometrical centrality dependence. We also observed that the $\alpha$ is far from the Gaussian case corresponds to $\alpha = 2$, and also it was found to be inconsistent with $\alpha \leq 0.5$, the conjectured value at the QCD critical point. We found a statistically significant decrease of the strength of the correlation function $\lambda(m_T)$ at low values of the transverse mass. We also observed that the normalized $\lambda/\lambda_{\max}$ function has similar behavior as $\lambda(m_T)$. However, this result is not consistent with predictions without in-medium $\eta'$ mass modification.

The author is thankful for the support of NKFIH grant K138136 and also K133046.

## References

1. W. Kittel, Acta Phys. Pol. B**32** (2001)
2. T. Csörgő, Heavy Ion Physics **15**, (2002)
3. R.H. Brown, R.Q. Twiss, Nature **177** (1956)
4. G. Goldhaber, S. Goldhaber, W. Lee, A. Pais, Phys. Rev. Lett. **3**: 181 (1959)
5. K. Adcox *et al*. [PHENIX Collaboration], Nucl. Inst. Methods Phys. Res., Sec. A **499** (2003)
6. P. Achard *et al*. [L3 Collaboration], Eur Phys. J. C **71** (2011)
7. A. Adare *et al*. [PHENIX Collaboration], Phys. Rev. C **97** (2018)
8. T. Csörgő, S. Hegyi, W. Zajc, Eur. Phys. J. C **36** (2004)
9. M. Csanád, Nucl. Phys. A. **774** (2006)
10. D. Kincses, Universe 4 **31** (2018)
11. S. Lökös, Universe 4 **11** (2018)
12. M. Csanád, S. Lökös, M. Nagy, Phys. Part. Nuclei **51** (2020)
13. D. Kincses, M. Nagy, M. Csanád, Phys. Rev. C **102** (2020)
14. B. Kurgyis, D. Kincses, M. Nagy, M. Csanád, https://arxiv.org/abs/2007.10173 (2023)
15. H. Rieger, Phys. Rev. B. **52** (1995)
16. T. Csörgő, S. Hegyi, T. Novák, W. Zajc, AIP. Conf. Proc. **828** (2006)
17. T. Csörgő, B. Lorstad, J. Zimányi, Z. Phys. C Part Fields **71** (1996)
18. J. Bolz, U. Ornik, M. Plumer, B.R. Schlei, Phys. Rev. D **47** (1993)
19. S.E. Vance, T. Csörgő, Phys. Rev. Lett. **81** (1998)
20. T. Csörgő, R. Vértesi, J. Sziklai, Phys. Rev. Lett. **105** (2010)
21. R. Vértesi, T. Csörgő, J. Sziklai, Phys. Rev. C **83** (2011)
22. R.D. Pisarski, F. Wilczek, Phys. Rev. D **29** (1984)
23. J.I. Kapusta, D. Kharzeev, Larry D. McLerran, Phys. Rev. D **53** (1996)
24. S. Weinberg, Phys. Rev. D **11** (1975)
25. Z. Huang, X.N. Wang, Phys. Rev. D **53** (1996)
26. Y. Kwon, S.H. Lee, K. Morita, Gy. Wolf, Phys. Rev. D **86** (2012)
27. D. Horvatic, D. Kekez, D. Klabučar, Phys. Rev. D **99** (2019)
28. H. Bekker *et al*. [NA44 Collaboration], Phys. Rev. Lett. **74** (1995)
29. H. Adhikary *et al*. [NA61/SHINE Collaboration] https://arxiv.org/abs/2302.04593 (2023)